\documentstyle{mn}
\input{psfig}
\title{NICMOS images of JVAS/CLASS gravitational lens systems}
\author[N. Jackson, E. Xanthopoulos, and I.W.A. Browne]
{N. Jackson, E. Xanthopoulos, and I.W.A. Browne\\
NRAL Jodrell Bank, University of Manchester, Macclesfield, Cheshire SK11 9DL}
\date{ }
\begin{document}
\maketitle

\begin{abstract}
We present Hubble Space Telescope (HST) infrared images of four gravitational lens systems 
from the JVAS/CLASS gravitational lens survey and compare
the new infrared HST pictures with previously published WFPC2 HST optical images and radio maps. 
Apart from the wealth of information that we get from the flux ratios and accurate positions 
and separations of the components of the lens systems that we can use as inputs for better 
constraints on the lens models we are able to discriminate between reddening and 
optical/radio microlensing as the possible cause of differences observed in the flux 
ratios of the components across the three wavelength bands. 
Substantial reddening has been known to be present 
in the lens system B1600+434 and has been further confirmed by the present infrared data.
In the two systems B0712+472 and B1030+074 microlensing has been pinpointed down as the main
cause of the flux ratio discrepancy both in the optical/infrared and in the radio, the radio 
possibly caused by the substructure revealed in the lensing galaxies.
In B0218+357 however the results are still not conclusive. If we are actually seeing the 
two ``true" components of the lens system then the flux ratio differences are attributed
to a combination of microlensing and reddening or alternatively due to some variability 
in at least one 
of the images. Otherwise the second ``true" component of B0218+357 maybe completely absorbed  
by a molecular cloud and the anomalous flux density ratios and large difference 
in separation between the optical/infrared and radio that we see 
can be explained by emission from either a foreground object or from part of the lensing galaxy.

\end{abstract}


\twocolumn

\section{Introduction}
Gravitational lens systems are important for a number of reasons.   
Individual lenses can be studied to determine a mass and information 
about the mass distribution of the lensing
galaxy or cluster. Moreover, detailed studies of the lensing galaxies
give clues to galaxy type and galaxy evolution (Keeton, Kochanek \&
Falco 1998). If the lensed object is variable, measurement of time
delays between variations of the multiple images, together with a good
mass model of the lensing object, 
allow one to estimate the Hubble constant (Kundi\'{c} et al.
1997; Schechter et al. 1997; Keeton \& Kochanek 1997; Biggs et al. 1998). Statistical
properties of lenses are also important. Both $q_{o}$ and $\lambda_{o}$ 
can be constrained by lens statistics, a good example being the recent
limit of $\lambda_{o} < 0.66$ for flat cosmologies  
(Kochanek 1996). For all these reasons, studies of larger samples of 
gravitational lenses are important.

The Jodrell Bank VLA Astrometric Survey (JVAS) and Cosmic Lens All Sky Survey (CLASS)
(Myers et al. 1998), 
in which 10,000 sources have so far been observed
with the VLA, is the most systematic radio survey so
far undertaken. It aims to look at all flat-spectrum radio sources in the
northern sky with flux densities $>$30mJy at 5 GHz. Flat-spectrum radio sources
are useful in this context for two main reasons; they have intrinsically
simple radio structures, making the effects of lensing easy to recognise, and  
they are variable, allowing time delays to be derived. The ease of
recognition means that this survey is statistically very clean and
should not be subject to any significant selection biases -- important  
in view of the results to be outlined below.

Here we present Hubble Space Telescope (HST) infrared imaging of four of
the JVAS/CLASS lens sample. The main aim of this work is to obtain
better constraints on the lensed images, in particular their flux ratios
and their positions relative to the lensing galaxy.
Saha \& Williams (1997) describe the importance of the use of the improved image positions 
and ratio of time delays between different pairs of images as a rigid constraint 
in the modelling of lenses especially when there is a limited number of observational
constraints. They note however, that lately it has become apparent (Witt \& Mao 1997) 
that in the case of quadruple-lens systems even well-determined image positions are not 
enough for accurate modelling, and thus they use non-parametric models.    

A second scope of the work is to compare the flux ratios of the components of the 
lens systems in the different wavebands (optical/near-infrared/radio). 
Variability of the source spectrum in connection with the time delay can lead to 
different flux ratios in different spectral bands. Another possibility is microlensing
in at least one image. The stellar mass objects can magnify only sufficiently 
compact sources so if the flux ratios in two spectral bands in two or more images of 
a multiply-imaged QSO are different, this can be a sign of microlensing. 
For example, in the gravitational lens system 0957+561, one suspects microlensing 
effects from the different flux ratios in the optical continuum light, broad-line flux 
and in the compact radio component. 
A third possibility is differential magnification, where 
parts of the source are more magnified than others. This can occur if the scale on 
which the magnification varies in the source plane is comparable with the size of the 
emitting regions; so the effect can be different for radiation of different wavelengths. 
The most spectacular effects of microlensing occur when the angular radius of the 
source is much smaller than that of the Einstein ring for the microlens (Refsdal \& Stabell 1997) 
but the effect has not been considered much up to now because of the long time scales that this occurs.
However, the time scale is no more important when comparing the flux ratios of the different 
images in multiply lensed quasars. 

Mao \& Schneider (1998) have suggested also that as in the case of microlensing for optical fluxes,
substructure in the lensing galaxy can distort the radio flux ratios and this can be another
cause of difference in the flux density ratios at different wavebands. 

Differential reddening can be an alternative explanation of differences observed in the 
flux density ratios of the lensed images at different wavebands.
In gravitational lens systems, especially those where the images are seen through the lens,
significant reddening by dust in the lens provides a natural explanation that must be 
considered first. 
There is considerable evidence for dust
reddening during passage through the lensing galaxy (Larkin et al. 1994, Lawrence,
Cohen \& Oke 1995; Jackson et al. 1998). Infrared pictures
are less subject to censorship by dust, generally giving the best view
of the lensed images, and enable the degree of extinction to be
quantified by comparison of brightnesses between images at different
wavelengths. Another advantage of infrared observations is that the
contrast between host galaxy starlight and AGN continuum is less than 
in the optical. Thus, for not very luminous lens
objects, we can expect to see arcs or rings formed by the extended
host galaxy starlight. The lens system B1938+666 (King et al. 1998) is a 
particularly spectacular example of this phenomenon. Our aim is also 
to study the colours and light distributions of the lensing galaxies
themselves. Keeton, Kochanek \&
Falco (1998) discuss measurements of optical HST data on a number of the
lensing galaxies, including B0218+357, B0712+472 and B1600+434, and use
optical observations and lens modelling to address the question of dark
matter distributions. Jackson et al. (1998) have published preliminary H-band
photometry of the JVAS/CLASS sample and show that all the lens systems
discovered so far have lensing galaxies of normal mass-to-light ratios
and that there is no evidence for dark galaxies in this sample.

In section 2 we give details the HST observations. Section 3 contains brief
descriptions and discussion of each individual object, and section 4
contains the conclusions.

\section{Observations}

All observations presented here were obtained with the Near Infrared
Camera/Multi-Object Spectrometer (NICMOS) on the HST. Observations were
taken with Camera 1, which has a pixel scale of 43~mas, and taken
through the F160W filter, which has a response approximately that of the
standard H-band. A list of observations, with exposure times and date of 
observations, is given in Table 1. Fits using the {\sc jmfit} program in
AIPS to the point spread functions generated by the TinyTim program 
(Krist 1997) yield full widths at half maximum (FWHM) of 
131\,mas for the NICMOS observations. The complementary Wide Field \&
Planetary Camera 2 data presented in the Figures have FWHM of 65\,mas
and 79\,mas for 555-nm and 814-nm images respectively.

\vskip 3mm 
\begin{table} 
\begin{tabular}{|c|ccccccc|} \hline 
Object & Obs date & Time & Observation number\\ \hline

B0218+357& 970817 & L & N44501KYM\\ 
         &        & S & N44501KZM\\ 
         &        & L & N44501L7M\\ 
         &        & S & N44501L8M\\ 
         &        & L & N44501LKM\\ 
         &        & S & N44051LLM\\ 
B0712+472& 970824 & L & N44503GAQ\\ 
         &        & S & N44503GBQ\\ 
         &        & L & N44503GKQ\\ 
         &        & S & N44503GLQ\\ 
B1030+074& 971122 & L & N44504VWQ\\
         &        & S & N44504VXQ\\
B1600+434& 970721 & L & N44505DYM\\ 
         &        & S & N44505DZM\\ \hline 
\end{tabular}
\caption{Log of the observations. L indicates an exposure time of 2048s
and S an exposure time of 576s.}
\end{table}

The data were processed by the standard NICMOS calibration pipeline at
STScI. 
Multiple exposures were combined by weighted  averaging  or medianing 
depending on whether we had two or more images available. Bad pixels and 
cosmic rays were rejected by using a rejection algorithm. The images are 
first scaled by statistics of the image pixels, here the "exposure" for
intensity scaling. The rejection algorithm used is the ``crreject" which 
rejects pixels above the average and works even with two images. The 
algorithm is appropriate for rejecting cosmic ray events.

As a second check we also added the images with binary image arithmetic 
within IRAF and divided the images by a number that is appropriate 
according to the exposures of the images and the number of the images 
involved so as to preserve the same counts/sec number as the initial 
NICMOS images. Cosmetically the result of both methods was the same.                   

A point spread function (PSF) was generated by the TinyTim programme
(Krist 1997) and deconvolved from the images using the {\sc allstar}
software, available within the {\sc NOAO IRAF} package. In cases where 
the point sources lie on top of extended emission, such as B0712+472, 
the PSF deconvolution was done by hand, shifting and scaling until the
smoothest residual map was obtained. Photometry and errors are based on the
quantity of PSF subtracted and the range of PSF subtraction which gives
a smooth residual map. In Fig. 1-4 we show the raw images and the
images with the point-sources subtracted.

\section{Results and discussion}

\begin{table*}
\begin{tabular}{|c|cccccc|} \hline
Survey & Name & \# images & $\Delta\theta "$ & z$_{l}$ & z$_{s}$ & lens galaxy \\ \hline 

JVAS & B0218+357 & 2 $+$ ring & 0.334 & 0.6847 & 0.96 & spiral \\
CLASS & B0712+472 & 4 & 1.27 & 0.406 & 1.34 & spiral \\
JVAS & B1030+074 & 2 & 1.56 & 0.599 & 1.535 & spiral \\
CLASS & B1600+434 & 2 & 1.39 & 0.414 & 1.589 & spiral \\ \hline 
\end{tabular}
\caption{The general characteristics of the four gravitational lens systems.}
\end{table*}

\begin{table*}
\begin{tabular}{ccccccccc} \hline
Object& Image & Flux density& Flux density& Flux density&
\multicolumn{4}{c}{Offset from A image}\\ 
 &&(555\,nm)&(814\,nm)&(1.6\,$\mu$m)&555\,nm&814\,nm&1.6\,$\mu$m&Radio\\ 
      &       &
\multicolumn{3}{c}{10$^{-20}$W\,m$^{-2}$nm$^{-1}$}&
\multicolumn{4}{c}{(RA,$\delta$,err) / mas}\\ \hline
B0218+357 & A & 2.0$\pm$0.5 & 2.0$\pm$0.5 & 12$\pm$1 &-&-&-&-\\
          & B & 13$\pm$1 & 19$\pm$2 & 19$\pm$1 &281,128,10&285,126,10&293,124,5&307,130\\
          & GAL & 6$\pm$2 & 13$\pm$2 & 15$\pm$2 & * & * & * & -\\
&&&&&&&&\\
B0712+472 & A  & 1.02$\pm$0.10 & 0.87$\pm$0.09 & 0.98$\pm$0.15 & - & - &
- & - \\
          & B  & 0.33$\pm$0.03 & 0.33$\pm$0.03 & 0.41$\pm$0.06 &66,$-$151,10&64,$-$149,10&*&57,$-$160\\
          & C  & 0.40$\pm$0.04 & 0.38$\pm$0.04 & 0.41$\pm$0.06 &812,$-$659,5&812,$-$660,5&804,$-$642,10&812,$-$667\\
          & D  & $<$0.02*       & 0.02* & 0.22$\pm$0.05 &*&*&1180,480,20&1163,460\\
          & GAL&  4.9$\pm$0.5   &  12$\pm$1 & 11$\pm$1          &814,157,3&801,153,3&789,168,3&-\\
&&&&&&&&\\
B1030+074 & A    & 26.94$\pm$0.10  & 27.37$\pm$0.07 & 47.6$\pm$0.4 &-&-&-&-\\
          & B    & 0.9$\pm$0.2 & 1.17$\pm$0.17 & 1.40$\pm$0.09 &930,$-$1256,4&931,$-$1247,4&918,$-$1244,5&935,$-$1258 \\
          & GAL  & 6? & 7.9$\pm$0.7 & 10.8$\pm$0.5 &882,$-$1155,10&878,$-$1143,10&864,$-$1162,5& \\
&&&&&&&&\\
B1600+434 & A  & 3.9$\pm$0.3 & 2.9$\pm$0.3 & 0.75$\pm$0.1 &-&-&-&-\\ 
          & B  & 2.0$\pm$0.2& 2.2$\pm$0.2&  0.53$\pm$0.1 &726,$-$1188,3&726,$-$1184,3&708,$-$1193,10&723,$-$1189\\
          & GAL &    *     & 9$\pm$2    &   4.5$\pm$1 &*&*&*&\\ \hline
\end{tabular}
\caption{Optical and F160W image photometry and positions. In each case
the galaxy photometry refers to the light within the Einstein radius,
apart from B0218+357 where the aperture was 3 square arcseconds. An
asterisk indicates an
image that is either invisible or impossible to deblend from another.
Plate scales of 45.5\,mas/pixel are assumed for the PC chip (555-nm and
814-nm images) and 43\,mas/pixel for NICMOS. More detailed discussion of
some of the optical images can be found in Jackson et al. (1998),
Koopmans, de Bruyn and Jackson (1998) and Xanthopoulos et al. (1998) for
B0712+472, B1600+434 and B1030+074 respectively. Errors in the radio
positions are 1\,mas or less. B1030+074A is approaching saturation on
the NICMOS image.}
\end{table*}

The general characteristics of all the four lenses can be found in Table 2.
Columns 1 and 2 present the name of the survey in which each lens system was
discovered and the name of the lens, while in columns 3, 4, 5, 6 and 7, one
can find the number of components in the lens system, the image separation in
arcsecs (the largest separation in the case of multiple components), the lens
redshift, the source redshift and the morphology of the lensing galaxy respectively.

\subsection{B0218+357}

The B0218+357 system consists of two images of a compact radio source
of a redshift 0.96 (Lawrence et al. 1995) separated by 334~mas. In addition
to the compact images there is a radio Einstein ring (Patnaik et al. 
1993). The lensing galaxy has a redshift of 0.6847 (Browne et al. 
1993). 
It has been detected in V, I and H-bands with the HST (Fig. 
1) and, as far as can be seen given the limits imposed by the signal in the
images, it has a smooth brightness
distribution and is roughly circular in appearance. The colours of the
galaxy are consistent  with its previous classification as a spiral
galaxy based on the presence of H1 absorption (Carilli, Rupen \& Yanny 1993),
strong molecular absorption (Wiklind \& Combes 1995) and large Faraday
rotations for the images. It is worth noting that the presence of a
rich interstellar medium  means that extinction along lines of sight
through the lensing galaxy may well be important, a possibility we will 
discuss 
below.

There are several anomalies associated with the optical properties
of the images in the B0218+357 system which might be sufficient to doubt
its classification as gravitational lens system, if it were not for the
existence of an Einstein ring (Patnaik et al. 1993), a measured
time delay (Corbett et al. 1995; Biggs et al. 1998) and high-resolution VLBA 15 GHz observations 
of B0218+357 (Patnaik et al. 1995). The observed anomalies are:
 
\begin{itemize}
 
\item The optical and infrared flux ratios (see Table 3 and Grundahl \&
Hjorth 1995) for the images are very different from the well established
radio ones (Patnaik et al. 1993).
\item The separation between the images may be less at optical and infrared
wavelengths than it is at radio wavelengths (see below).
\item The optical/infrared colours of the two images are very different.
\end{itemize}

The V and I flux densities of B are consistent with the ground-based
optical spectrum (Browne et al. 1993, Stickel et al. 1996; Lawrence et
al. 1995). Similarly the H flux density is what one might expect if B
has a spectrum typical of a BL Lac-like object. The A/B flux density
ratios are, however, unexpected. The most obvious result is that they
are very different from the 3.7:1 measured in the radio (Patnaik et
al., 1993; Biggs et al., 1998). This could arise from micro-lensing of
the optical/infrared emission and/or from extinction. However,
extinction following a normal Galactic reddening law (Howarth 1983)
is not consistent
with the optical/infrared colours. This is because, if we attribute
everything to extinction, the necessary amount of reddening to give a
1:7 ratio in the V-band image should give about 1:2 in I, assuming
an intrinsic 3.7:1 ratio of the radio images. Even if we attribute the
gross difference between the radio and optical/infrared ratios to
micro-lensing, the fact that the V and I flux densities of image A are
the same is not consistent with the A emission being a reddened
version of the image B. Furthermore, if the flux densities for image B
listed in Table 3 were severely contaminated with lensing galaxy
emission, this still would not help. One last possibility would be to
invoke some variability in at least one image, although this would not
explain the problems arising from comparison between the V and I images 
which were taken within an hour of each other.

Hjorth (1997) has suggested that the separation between the images may
be less at optical and infrared wavelengths (i.e. $\sim$300~mas) than
it is at radio wavelengths (i.e. 334~mas). It is important to
establish if this difference in image separations is real since, if
true, it would immediately indicate that we are not seeing two images
of the same object in the optical/infrared band. 

The radio image separation is well established to be 334$\pm$1 mas (Patnaik et
al. 1993; Patnaik et al. 1995). The separations (and flux densities)
given in Table 3 are the result of by-eye fitting of point spread
functions to the HST pictures in such a way as to obtain smoothest
residuals.  The optical pictures have much lower signal-to-noise than
the infrared ones, particularly for the A component and have
correspondingly bigger errors. The image separation derived from
fitting to the NICMOS picture is 318$\pm$5~mas, a value much smaller than the
radio separation given above. This H-band separation is close to the one we find 
from the optical V and I WFPC2 images (308$\pm$10~mas and 311$\pm$10~mas in 
V and I respectively) but slightly larger than the optical separation found by Hjorth (1997) 
(296$\pm$10~mas and 299$\pm$10~mas from V and I respectively).  
The method we employ does not explicitly take account of the
light from the lensing galaxy. 

Models consisting of an exponential
disk plus point images have been fitted to the data by McLeod et
al. (1999). We have tried the same approach. 
A point spread function (PSF) was generated by the TinyTim programme
(Krist 1997) and was fitted simultaneously to the A and B components 
of the lens system. We have tried fitting and subtracting the PSF for different
positions of the two components. By attributing some of the emission around 
the B image (about 30\%) to the core of the lensing galaxy (central 0.4 $\times$ 0.4 
arcsec) we increase the separation until no flux is left (by subtracting both 
the components and the lensing galaxy) and so we find the maximum best fit 
to give a separation of 335 mas between the two components. The lensing galaxy 
has then an offset of 0.54,0.34 x, y pixels from the B component or it is 
27 mas away at a  position angle of 260$^\circ$ from B. By taking into account the 
light contribution from the lensing galaxy the method removes the 
discrepancy between the radio and optical/infrared separation.  
A limit can be deduced for the fraction of the
light from the region of the B image which is contributed by the
lensing galaxy. If we examine the optical spectra obtained at a number
of different epochs (e.g. Browne et al. 1993, Stickel et al. 1996;
Lawrence et al. 1995) they show no significant 4000\AA\ break nor any
G-band absorption feature found in nearly all galaxy types (Bica \&
Alloin 1987). Of the light in the spectra $\leq$15\% comes from the
lensing galaxy. This implies that around the region of the B image, at
an observed wavelength of $\sim$670~nm the galaxy contributes no more
than $\sim$30\% of the total flux density, allowing for slit losses
and the fact that the A image makes a small contribution to the total
light. Thus, if the lensing galaxy is as blue or bluer than the BL
Lac-like spectrum of lensed object, the model with a 30\% contribution
from the galaxy at 1.6 $\mu$m is only just consistent with the
spectral data. Hence we conclude that, though {\it a priori} one would
expect the radio and optical/infrared image separations to be the
same, there is some evidence to suggest that they are different. The
anomalous flux density ratios may also in some way be related to the
separation problem.

So as mentioned above even if one takes into consideration the contribution 
of the light of the lensing galaxy to the B component there still is some 
doubt that the separation problem between the optical/infrared and radio
is completely resolved.  

Hjorth (1997) find from their optical data that the bright images that
they observed could be identified with the radio A and B components
because of the excellent agreement between the position angles, which is also in
agreement with Grundahl \& Hjorth (1995). They attribute the shorter
separation to the extendedness of component A and they argue that what
we see is what is not covered by the molecular cloud (Grundahl \& Hjorth 1995;
Wiklind \& Combes 1995). However a potential problem with this interpretation 
is that the bright image appears to be a point source and not extended as 
would be expected in this case.  
The alternative explanation that they offer, that A is actually the core of 
the lensing galaxy while the counterpart to the B image is somehow swamped in 
the light of the galaxy core needs also to explain the fact that since B is a 
BLLac as proved from both ground based spectra and its V, I  and H fluxes and 
the main dominant source of the optical continuum we are expecting to see a 
radio source at the location of the lens BL Lac (Urry \& Padovani 1995) which 
we do not.   
Grundahl \& Hjorth (1995) actually note that
it is possible (due to the small intrinsic extent of A - 1 mas at high
radio frequencies, so 5 pc at the redshift of the galaxy) that the
molecular cloud can cover the entire image A.
If this is the case then we are left with the difficult question concerning the nature of the
A emission (hereinafter we refer to the optical/infrared emission near A as A$^{*}$) 
if it is not an image of the AGN.
The fact that the
true A image is obscured does not mean that all the light from host
galaxy of the lensed object must be hidden too; some of its emission
could be lensed  and give  rise to A$^{*}$. A difficulty
with this idea is that the A$^{*}$ image is compact and of high
surface brightness not extended as one might expect if it were an
image of some part of the AGN host galaxy.  Also there is no sign of
an equivalent (B$^{*}$) image near B.  

An alternative possibility is that A$^{*}$ is not a gravitational
image at all but is part of the emission from the lensing galaxy (or
even a foreground object). In this scenario one has to attribute the
proximity of A$^{*}$ to A, and the fact that the A$^{*}$ lies at the
expected position angle for a lensed image, to coincidence.

In the end, we find none of the possibilities satisfactory. It could be that we are 
deluted about the reality of separation difference and there is nothing to 
explain but the different colours of the A and B images.

\begin{figure*}
\psfig{figure=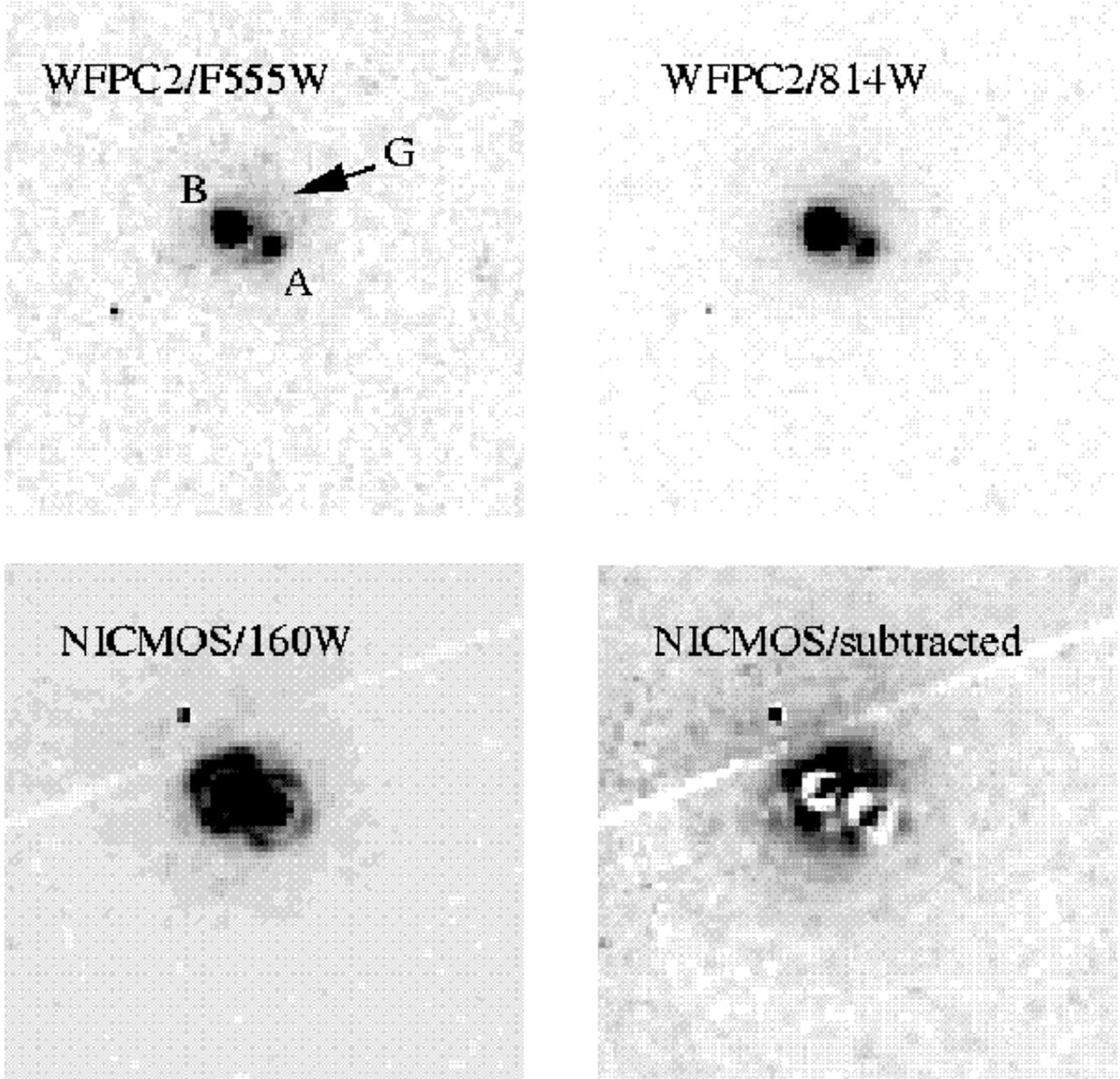,width=18cm,angle=0}
\caption{Top: HST optical (555-nm and 814-nm) images of B0218+357.
Bottom: HST infrared (1.6\,$\mu$m) images. North is up and East to the left. Each image is 
3\farcs 69 on a side. For the WFPC2 images the scale is 45.5 mas/pixel and for the NICMOS 
images 43 mas/pixel.}
\end{figure*}

\subsection{B0712+472}

Radio observations revealed this system to be a four-image gravitational
lens, although only the A, B and C images were visible in the optical
WFPC2 images (Jackson et al. 1998). In the new NICMOS image (Fig. 2) we
clearly see the D image also. We use the new data to revisit the
question of the anomalous flux density ratios discussed in the earlier work.

In the radio, optical and infrared bands, images A and C have the same
flux density ratio (about 2.5:1) within the errors. The major discrepancy
concerns the flux densities of B and D. 
The invisibility of D in V band
and its marginal visibility in I can be ascribed to reddening in the
lensing galaxy. 
Reddening was also initially thought to be the case for the large difference 
observed in the radio and optical flux density of component B (it has 80-90\% of the flux of 
component A in the radio and only 30\% of A in 555-nm). 
However, the fact that the B/A ratio remains constant throughout the optical and
infrared within the errors while the inferred reddening at 555\,nm, $A_{V}\sim$1 mag,
requires the reddening to fall to $<0.2$ mag at 1.6\,$\mu$m which is clearly not the case here
as seen from the NICMOS images, argues against it. 
Variability is now also unlikely, since the infrared
and optical observations were taken over a year apart, much longer than
the likely time delay between images, and yet the B/A optical/infrared 
flux density ratio has remained relatively constant (and very different from the
radio flux density ratio). It seems therefore likely that we are, as suggested
by Jackson et al. (1998), seeing an episode of microlensing. 
Further
monitoring of this system over periods of years or tens of years (the
typical timescale of microlensing in such a system) should
reveal a gradual increase in the optical and infrared B/A flux density ratio.

The positions of all objects are (just) consistent between the optical,
infrared and radio images. The exception is an apparent shift of the
position of the lensing galaxy at different wavelengths which is
small but systematic with increasing wavelength. 

\begin{figure*}
\psfig{figure=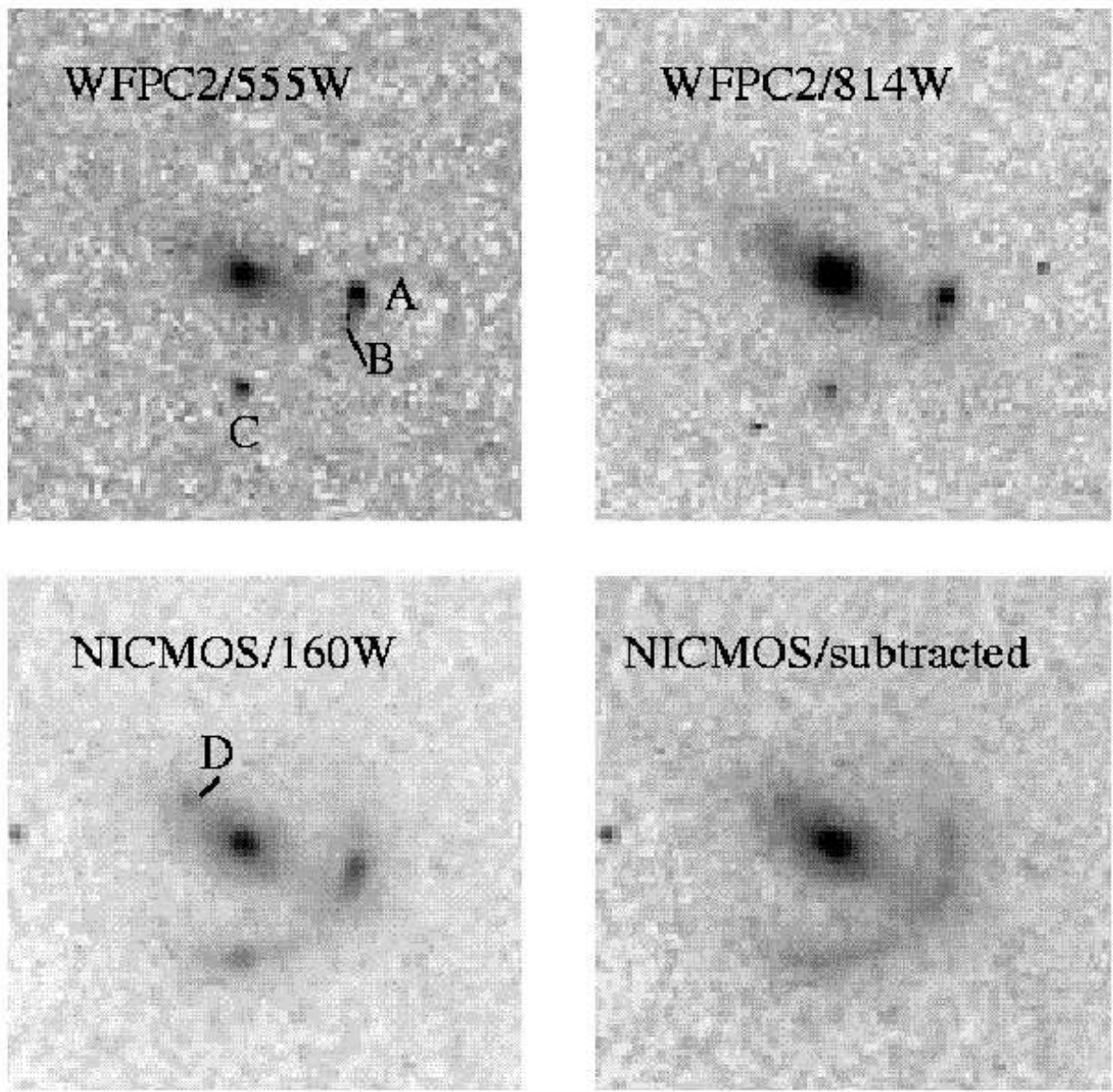,width=18cm,angle=0}
\caption{Top: HST optical (555-nm and 814-nm) images of B0712+472.
Bottom: HST infrared (1.6-$\mu$m) images, without and with subtraction
of the point sources A,B,C,D. Note the arc that remains in the infrared
NICMOS picture after subtraction of the point sources. Each image is 
3\farcs 69 on a side. North is up and East to the left.
For the WFPC2 images the scale is 45.5 mas/pixel and for the NICMOS 
images 43 mas/pixel.}
\end{figure*}

The  lensed object appears to have a spectral index of $\alpha\sim 0$, 
where the flux F($_{\lambda}$)$\propto\lambda^{\alpha}$ in the rest-frame
wavelength range of 240\,nm -- 690\,nm. This is much redder than the
typical values of $\alpha\sim -1$ found for quasars (e.g. Jackson \& 
Browne 1991), implying that little quasar continuum is present. It has already
been remarked that the object is severely underluminous (Jackson et al. 1998);
it now appears that the continuum is unusually red for a quasar and is
thus likely to have a substantial contribution from starlight, although
some AGN component must be present to give the broad lines. 
This reinforces the conclusion that we may be dealing with an example  
of a mini-quasar inhabiting a host which is bright compared to the AGN
component. Given the presumed
magnifications (Jackson et al. 1998), the intrinsic radio luminosity of  
log($L_{\rm 5GHz}$/W\,Hz$^{-1}$sr$^{-1}$)$\simeq$24.4 is only just in the 
radio-loud category as defined by Miller, Peacock \& Mead (1990). 
The host galaxy of
the lensed object is also visible, smeared into an arc in the NICMOS
picture (Fig. 2) and is considerably redder than the quasar at its
centre; the arc is invisible in the optical pictures.
 
The optical-infrared colours of the lensing galaxy 
are consistent with the conclusion by Fassnacht \& Cohen (1998) that the
lens is an early-type galaxy. The 
4000\AA\ break, at the galaxy's redshift of 0.41 (Fassnacht \& Cohen
1998) occurs between 555\,nm and 814\,nm.

\subsection{B1030+074}

The discovery of the lensed system B1030+074 was reported by
Xanthopoulos et al. (1998). The lensed images are of a quasar/BL Lac of
redshift 1.535 (Fassnacht \& Cohen 1998) and are separated by
1.56~arcsec with a radio flux density ratio in the range 12 to 19. The
lensing galaxy has a redshift of 0.599 and its spectrum is typical of
an early type galaxy (Fassnacht \& Cohen 1998).

In Fig. 3 we show the WFPC2 V and I pictures (Xanthopoulos et al.,
1998), together with our new NICMOS 1.6$\mu$ picture. In all three
bands the lensing galaxy is seen near the fainter B image. The peak of
the galaxy light lies on the line joining the two images but there is 
also a secondary emission feature to the SE of the main part of the 
galaxy. Xanthopoulos et al. suggest that this secondary  peak may be 
a spiral arm. The H-band data indicate that the colour of this
feature is similar to that of the rest of the lensing galaxy, possibly
suggesting that it is not a spiral arm but either part of the main galaxy or
a companion object.

The colours of the lensed images are consistent with our knowledge
that the lensed object is a quasar or BL Lac object. Of interest,
however, are the optical/infrared flux density ratios of the images
which are considerably higher than any of those measured at radio
frequencies. It is tempting to attribute such differences to
extinction but the fact that the H-band ratio is even larger than V
and I ratios argues against this. We suggest that the differences are
best explained as arising from microlensing. Substructure in the lensing 
galaxy can also distort the radio flux ratios (as in microlensing for optical fluxes),
in particular for highly 
magnified images, without appreciably changing image positions. As Mao \& Schneider
(1998) note such substructure can be for example spiral arms in disk galaxies or 
structure caused by continuous merging and accretion of subclumps, as may possibly be   
the case for B1030+074, and can have a little effect on time delays and the determination
of H$_{0}$.

\begin{figure*}
\psfig{figure=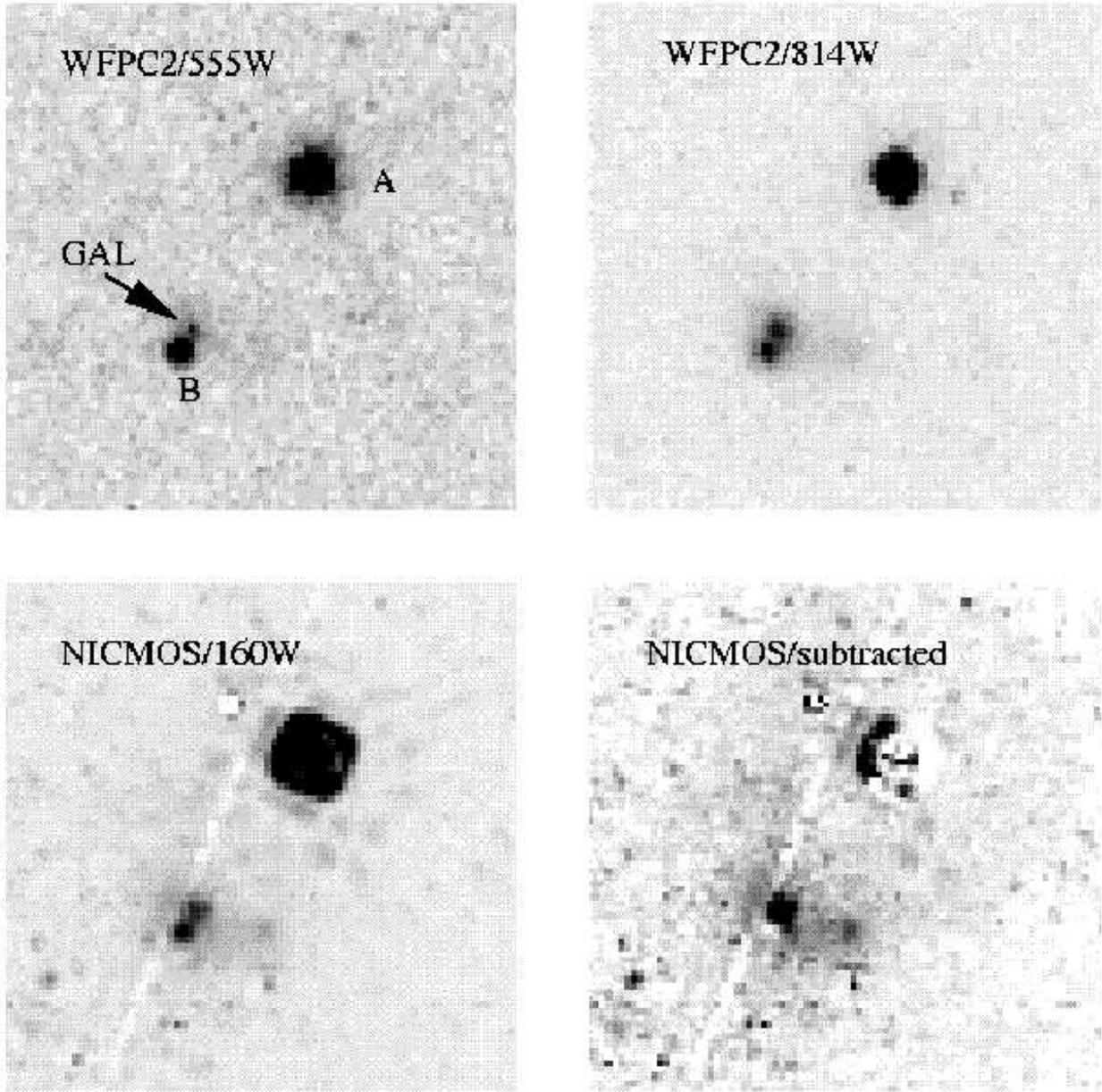,width=18cm,angle=0}
\caption{Top: HST optical (555-nm and 814-nm) images of B1030+074.
Bottom: HST infrared (1.6-$\mu$m) images, raw (left) and with the two
point objects A and B subtracted (right). The very bright A image makes
a good PSF subtraction difficult. Each image is 3\farcs 69 on a side. 
North is up and East to the left. 
For the WFPC2 images the scale is 45.5 mas/pixel and for the NICMOS 
images 43 mas/pixel.}
\end{figure*}

\subsection{B1600+434}

The lens system B1600+434 was discovered in the first phase of the CLASS
survey (Jackson et al. 1995). It is a two-image system of separation
1392\,mas; the lensing galaxy lies close to the southeastern component.
In Fig. 4 we show the HST 555-nm and 814-nm images, together with the
new NICMOS 1.6-$\mu$m image. The lensing galaxy is an edge-on spiral
(Jaunsen \& Hjorth 1997; Kochanek et al. 1999),
which has been modelled in detail by Maller, Flores \& Primack (1997)
and by Koopmans et al. (1998).

Photometry of this object is affected by the coincidence of the
southeastern image, B, with the lensing galaxy. It is also likely that
both optical images are variable (Jaunsen \& Hjorth 1997). Jaunsen \&
Hjorth (1997) present ground-based BVRI photometry from which they
deduce substantial reddening to be present in image B, almost certainly
due to passage through the lensing galaxy. The infrared data lend
support to this view, as the flux density 
ratios of the images in the infrared and radio
are indistinguishable to within the errors. 

The unreddened image A has a spectral index $\alpha\sim -1.7$, which is
roughly the spectral index of a normal quasar.

\begin{figure*}
\psfig{figure=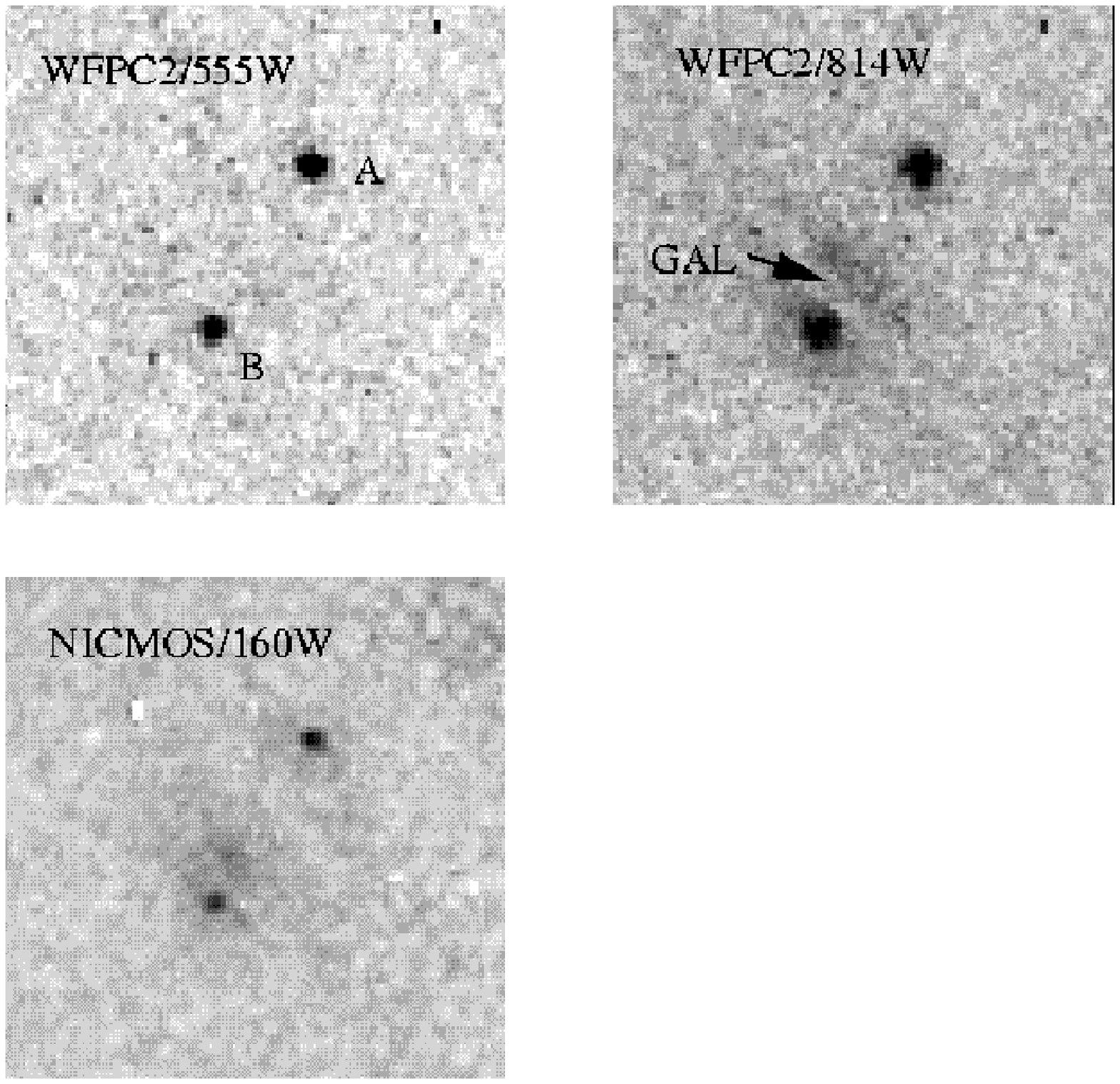,width=18cm,angle=0}
\caption{Top: HST optical (555-nm and 814-nm) images of B1600+434.
Bottom: HST infrared (1.6-$\mu$m) image. Each image is 3\farcs 69 on a side.
North is up and East to the left. 
For the WFPC2 images the scale is 45.5 mas/pixel and for the NICMOS 
images 43 mas/pixel.}
\end{figure*}

\section{Summary and conclusions}
 
Our NICMOS observations have proved successful at detecting both lensing
galaxies and lensed objects. We find that often the flux density ratios
of
the images are different from those measured at radio wavelengths. Having
infrared colours enables us to distinguish between extinction and
microlensing as explanations for the different flux density ratios. We find
evidence for both in different systems. Reddening is clearly
affecting the B image in B1600+434 and may be part of the explanation for the
puzzling system B0218+357. On the other hand, in both B0712+472 and
B1030+074, the fact that the infrared and optical image flux density 
ratios are the same but different from the radio ones is strong evidence that
microlensing is important in these objects. 
 
On a larger scale microlensing can have an important effect on lensing statistics
(see Bartelmann \& Schneider 1990) and the fact the we see evidence in maybe 
three of the four systems examined here argues that the phenomenon is not as 
rare as previously thought. If microlensing is considered, single highly-magnified 
images can occur and for example the tight correlation between total magnification 
and flux ratio is weakened by microlensing. 
 
The arcs seen in B0712+472 illustrate another important fact. Many of
the lensed objects have relatively subluminous AGN and in some cases the
infrared emission from the host galaxy starlight dominates over that
from the AGN. This is when arcs are seen. Other examples are B1938+666 (King
et al. 1998), B2045+265 (Fassnacht et al. 1999) and B1933+503 (Marlow et
al. 1999). With an arc the lensing galaxy potential is probed at many
points and provides useful additional constraints for the lens mass
models.

\section*{Acknowledgments}

This research was based on observations with the Hubble Space Telescope,
obtained at the Space Telescope Science Institute, which is operated by
Associated Universities for Research in Astronomy, Inc., under NASA
contract NAS5-26555. This research was
supported by the European Commission, TMR Programme, Research
Network Contract ERBFMRXCT96-0034 ``CERES''. We thank C. Kochanek for
comments leading to an improvement in the first version of the paper.

\end{document}